\pdfoutput=1
\PassOptionsToPackage{table}{xcolor}
\PassOptionsToPackage{final}{hyperref}
\documentclass[nonacm,screen,table]{acmart}

\makeatletter % This code section allows cdashline inside a table to work properly
\def\adl@drawiv#1#2#3{%
        \hskip.5\tabcolsep
        \xleaders#3{#2.5\@tempdimb #1{1}#2.5\@tempdimb}%
                #2\z@ plus1fil minus1fil\relax
        \hskip.5\tabcolsep}
\newcommand{\cdashlinelr}[1]{%
  \noalign{\vskip\aboverulesep
           \global\let\@dashdrawstore\adl@draw
           \global\let\adl@draw\adl@drawiv}
  \cdashline{#1}
  \noalign{\global\let\adl@draw\@dashdrawstore
           \vskip\belowrulesep}}
\makeatother

\usepackage{tabularx}
\usepackage{makecell}
\usepackage{hyperref}

\newcommand\tablefootnote[1]{% Makes footnote in a table unclickable
  \begin{NoHyper}%
  \footnote{#1}
  \end{NoHyper}%
}
\newcommand\tablecref[1]{% Makes cref in a table unclickable
  \begin{NoHyper}%
  \cref{#1}
  \end{NoHyper}%
}

\AtBeginDocument{%
  \providecommand\BibTeX{{%
    \normalfont B\kern-0.5em{\scshape i\kern-0.25em b}\kern-0.8em\TeX}}}

\setcopyright{acmcopyright}
\copyrightyear{XXX}
\acmYear{XXX}
\acmDOI{XXXXXXX.XXXXXXX}

\acmPrice{15.00}
\acmISBN{978-1-4503-XXXX-X/18/06}

\newcommand{\ket}[1]{\vert #1 \rangle}
\newcommand{\ketbra}[1]{\vert #1 \rangle \langle #1 \vert}

\newcommand{\tensor}{\otimes}

\usepackage[capitalise]{cleveref}
\crefformat{footnote}{#2\footnotemark[#1]#3}

\usepackage{arydshln}
\usepackage{booktabs}
\usepackage{footnote}
\usepackage[table]{xcolor}    % loads also »colortbl«

\usepackage{ifdraft}
\newcommand{\authnote}[3]{}
\newcommand{\onote}[1]{\authnote{Or}{#1}{blue}}
\newcommand{\snote}[1]{\authnote{Shai}{#1}{purple}}
\begin{document}

\title{Uncloneable Cryptography}

%\author{Or Sattath}
%\email{sattath@bgu.ac.il}
\author{Or Sattath}
\authornotemark[1]
\email{sattath@bgu.ac.il}
\orcid{0000-0001-7567-3822}
\affiliation{%
    \institution{Ben-Gurion University of the Negev}
    \city{Beersheba}
    \country{Israel}
}


\begin{abstract}
The no-cloning theorem asserts that, unlike classical information, quantum information cannot be copied.
This seemingly undesirable phenomenon is harnessed in quantum cryptography. Uncloneable cryptography studies settings in which the impossibility of copying is a desired property, and achieves forms of security that are classically unattainable.
The first example discovered and analyzed was in the context of cash. On the one hand, we want users to hold the cash; on the other hand, the cash should be hard to counterfeit. Quantum money uses variants of the no-cloning theorem to make counterfeiting impossible.

In the past decade, this field developed in various directions: several flavors of quantum money, such as classically verifiable, locally verifiable, semi-quantum, quantum coins, and quantum lightning were constructed. New uncloneable primitives were introduced, such as uncloneable signatures, quantum copy protection for classical software, pseudorandom states, and several uncloneable forms of encryption. This work is a gentle introduction to these topics.
\end{abstract}

\ccsdesc[500]{Security and privacy~Cryptography}
\ccsdesc[500]{Theory of computation~Quantum computation theory}
\ccsdesc[300]{Social and professional topics~Intellectual property}

\keywords{Quantum Cryptography, No-cloning}

\maketitle
\begin{flushright}
 In memory of Stephen Wiesner, 1942–-2021.
\end{flushright}

\onote{Todo: 
\begin{enumerate}
        \item 
\end{enumerate}
}

Intractable computational problems are a barrier for algorithm designers. 
Cryptographers are modern lemonade makers. Their lemons are these intractable problems, which they squeeze into sweet lemonade: secure cryptographic protocols.
Why is a lemon even required? Because it lets us assume there is something an adversary \emph{cannot} do. 
Intractable problems can give the honest user an advantage: for example, the honest user can multiply two large primes. The honest user knows the prime factors of the resulting number; yet, it is widely believed that a classical adversary cannot (efficiently) find these factors.  

% Cryptographers have been squeezing this computational intractability lemon since the 70s. Are there any other lemons on which cryptography could be based?
% Quantum mechanics has quite a few peculiarities.
% One notable example is the no-cloning theorem, which states that quantum information cannot be cloned.
% This is in sharp contrast with classical information, which can be copied easily. Additionally, the act of measuring a state collapses it.
% This transformation is irreversible: there is no way to ``rewind'' this process and return to the pre-measured state.
%Lastly, non-local games have the property that classical strategies get a ``low score'' compared to the optimal quantum strategies.

% Indeed, these ``no-go'' results turn out to be the new lemons that are the main horse-power in quantum cryptography. 
% Uncloneable cryptography---the main focus of this review---combines the no-cloning lemon with other such lemons (mostly the computational intractability and no-rewinding lemons). See \cref{fig:lemons}. 

\begin{figure*}
    \centering
    \includegraphics[width=0.9\textwidth]{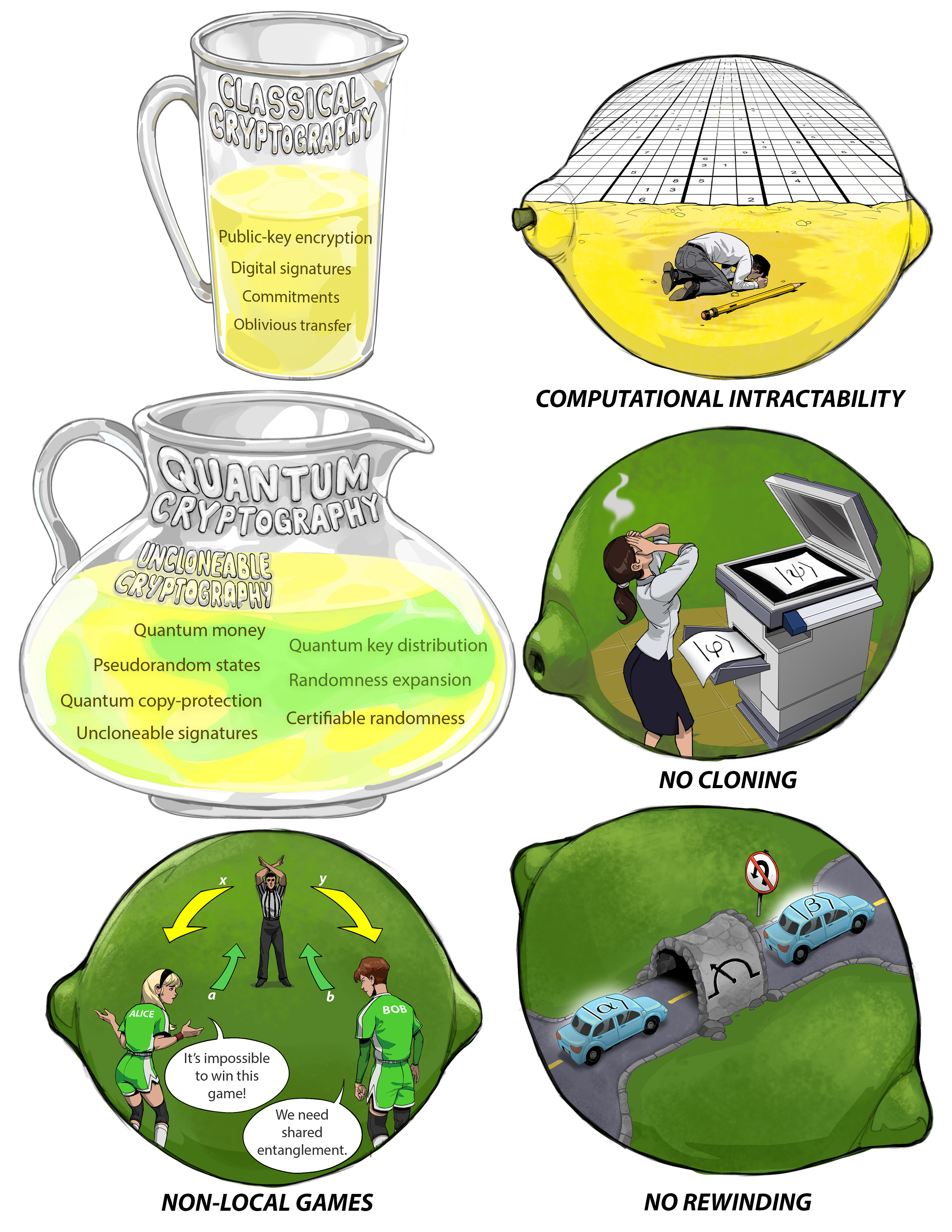}
    %\caption{Classical and quantum cryptography.}
    \caption{%Classical and quantum cryptography. 
    Classical cryptography relies on the computational intractability lemon. Uncloneable cryptography combines the computational intractability lemon, with the no-cloning lemon---see more details in the main text. 
    %Quantum cryptography, which contains uncloneable cryptography as a subfield, uses two other fresh (hence, the green hue) lemons: the act of measuring a state collapses it.
    %Quantum cryptography contains uncloneable cryptography as a subfield. 
    For the reader's orientation, we mention that uncloneable cryptography is a sub-field of quantum cryptography.
    Quantum cryptography uses two other no-go phenomena as fresh (hence, the green hue in the illustration) lemons: the act of measuring a state collapses it, and this transformation is irreversible: there is no way to ``rewind'' this process and return to the pre-measured state.
    Additionally, non-local games have the property that classical strategies get a ``low score'' compared to the optimal quantum strategies. This review focuses on uncloneable cryptography, and therefore no rewinding and no cloning are not discussed in detail. Illustration credits: Vince Serrano. }
    \label{fig:lemons}
\end{figure*}

Cryptographers have been squeezing this computational intractability lemon since the 70s. Are there any other lemons on which cryptography could be based?
Quantum mechanics has quite a few peculiarities.
One notable example is the no-cloning theorem, which states that quantum information cannot be cloned. Uncloneable cryptography---the main focus of this review---uses the no-cloning lemon as its main ingredient. For a broader perspective, see \cref{fig:lemons}. 
%Uncloneable cryptography is a subfield of quantum cryptography, uses two other fresh lemons
%Additionally, the act of measuring a state collapses it.
%This transformation is irreversible: there is no way to ``rewind'' this process and return to the pre-measured state.
%Lastly, non-local games have the property that classical strategies get a ``low score'' compared to the optimal quantum strategies.

%Indeed, these ``no-go'' results turn out to be the new lemons that are the main horse-power in quantum cryptography. 

% No cloning theorem. The statement itself. Resemblance to no-fresh-sampling.

% Irrevarsability of quantum measurement. 

% Wisener's scheme. 

% Variants of quantum money. 

% Implementations. Challenges. 

% Pseudorandom states and pesudorandom unitaries.

%\subsection*{No-cloning}  
\subsection*{Quantum states and no-cloning}  
%\paragraph{No-cloning}

Suppose you are given a sample from a probability distribution you know nothing about. Can you create another independent sample from that distribution? Clearly, the answer is no.
 The no-cloning theorem states that quantum information behaves the same way:  if you are given an unknown quantum state $\ket{\psi}$, it is impossible to create two identical copies of it. Also, the proofs of these two statements are very similar. The no-cloning result and its simple proof use quantum jargon, which can be safely skipped for those unfamiliar. 
\begin{theorem}[\cite{WZ82,Die82,Par70}]
The transformation $\ket{\psi}\mapsto \ket{\psi}\tensor\ket{\psi}$ is non-linear, and therefore is not permissible by quantum mechanics.
\end{theorem}
A qubit---the most basic unit of quantum information---is a linear combination of the "0" and "1" state of a single bit, and therefore is analogous to a bit. Yet, the theorem above shows that this analogy shouldn't be taken too literally: unlike bits, which can be copied, there are aspects in which a qubit resembles a (classical) sample from a distribution, since in both cases one copy is not enough to create two independent copies.
\begin{proof}
By definition, 
\begin{align}
\label{eq:cloning_0}\ket{0}&\mapsto \ket{0}\tensor\ket{0} \\
\label{eq:cloning_1}\ket{1}&\mapsto \ket{1}\tensor\ket{1} \\
\label{eq:cloning_plus}\ket{+}&\mapsto \ket{+}\tensor\ket{+}=\frac{1}{2}(\ket{0}\tensor\ket{0}+\ket{0}\tensor\ket{1}+\ket{1}\tensor\ket{0}+\ket{1}\tensor\ket{1})    
\end{align}
By summing \cref{eq:cloning_0,eq:cloning_1}, and assuming that the transformation is linear, we have $\ket{+}:= \frac{1}{\sqrt 2} (\ket{0}+\ket{1})\mapsto \frac{1}{\sqrt 2}(\ket{0}\tensor \ket{0}+\ket{1}\tensor\ket{1})$, which contradicts \cref{eq:cloning_plus}.
\end{proof}
Note that this theorem only shows that it is impossible to create \emph{two} \emph{exact} clones of \emph{all} states given a \emph{single} copy. 
There are many no-cloning variants where these restrictions are relaxed. 
%For example,  Bruß et al.~\cite{BDE+98} discuss optimal cloner that works when given one of two (non-orthogonal) states $\ket{\alpha},\ket{\beta}\in \mathcal C^2$, the goal is to create two states with the highest possible fidelity to two clones. 
For example, Bruß et al.~\cite{BDE+98} discuss optimal \emph{state dependent} cloners. Here, the cloner (C) receives one of two predetermined states, $\ket{\alpha},\ket{\beta}\in \mathbb{C}^d$; The cloner's goal is to have $F(C(\ket{\alpha}),\ket{\alpha}\tensor \ket{\alpha})\geq x$ and $F(C(\ket{\beta}),\ket{\beta}\tensor \ket{\beta})\geq x$ for $x$ as large as possible, where $F$ is a quantity called the \emph{fidelity} which measures how close two given states are. 
Werner~\cite{Wer98} discusses an optimal $N\rightarrow M$ cloner for any $M>N$: the cloner receives $N$ copies of a Haar random state $\ket{\psi}\in \mathbb C^d$ and outputs $M$ registers which have the highest possible fidelity with $\ket{\psi}^{\tensor M}$.
Among other things, Werner's result shows that $poly(n)$ copies of a Haar random state on $n$ qubits would yield an exponentially small fidelity when trying to produce (only) one additional clone.

%\section*{Section} and some text
%\subsection*{SubSection} and some text
%\subsubsection*{Subsubsection} and some text

\subsection*{Quantum money and uncloneable signatures}
%\paragraph{Quantum money and uncloneable signatures}
The first uncloneable primitive, and arguably the first work in quantum cryptography, was Wiesner's private quantum money scheme~\cite{Wie83}.\footnote{The first version of this manuscript was submitted circa 1969.} Wiesner's motivation was to design ``money that it is physically
impossible to counterfeit''.
His construction is as follows: Upon minting, the bank creates a banknote using $n$ qubits, where each qubit can be in one of the  4 states $\ket{0},\ket{1},\ket{+}:= \frac{1}{\sqrt 2}(\ket{0}+\ket{1}),\ket{-}:= \frac{1}{\sqrt 2}(\ket{0}-\ket{1})$. The banknote also contains a serial number. For example, the $9$th banknote could be $(\ket{1}\tensor\ket{-}\tensor\ket{+}\tensor\ket{-}\tensor\ket{0},9)$ (here $n=5$, which is too small in practice). The bank also maintains a database, containing a classical description of the quantum state for each of the serial numbers. Money can be verified in each of the bank branches: each bank branch needs to have a copy of the database mentioned above. 
To verify the $9$th banknote in the previous example, the bank would measure the first qubit of the proclaimed state in the 0/1 basis, and reject if the outcome is not 1 (recall that the first qubit is supposed to be $\ket{1}$, and therefore measuring it in the standard basis should return 1). Otherwise, it will verify the second qubit by measuring it in the +/- basis,  and rejecting if the outcome is not~-. This is repeated for all the $n$ qubits. 

There are variations of this scheme that are noise-tolerant (i.e., money passes verification even if some constant fraction of the qubits are disturbed arbitrarily). A complete analysis proving the security of Wiesner's scheme appeared roughly 40 years later~\cite{PYJ+12,MVW12}.

Since its inception, quantum money has been improved across multiple dimensions and properties, which are discussed in the rest of this section.
\cref{tab:quantum_money} contains a cell for every combination of these properties. 
Every such cell contains representative constructions and the common term used for that set of properties, which might not be so obvious: for example, a particular type of keyless quantum money is called quantum \emph{lightning}.
%Every such cell contains representative constructions and the common term used for that set of properties, which might not be so obvious: for example, schemes in which the money states are indistinguishable copies are called quantum \emph{coins}.
Note that in some cases, one notion is stronger than another. For example, public quantum money, which will be discussed next, is stronger than private quantum money\footnote{stronger in the sense that every public quantum money scheme is also a private quantum money scheme, but the converse does not hold}. 
In many cases, a construction for some stronger variant is known, while a construction that achieves \emph{only} the weaker one is not known. We emphasize that one could try to achieve the weaker variant, while improving upon the properties which do not appear in the table, such as, weakening the underlying assumptions, removing the need for an oracle, making the scheme noise-tolerant, improving the efficiency, etc. 

\paragraph{Types of verification keys.}
Wiesner's money is reminiscent of credit cards, where a trusted party is involved in any transaction. This type of quantum money is called \emph{private} since the bank's private key is needed to verify the money. There are three other types of verification keys, shown in the vertical axis in the table, which are discussed next. 

Public quantum money, also known as locally verifiable quantum money, resembles cash: verification can be done without the bank, using the bank's \emph{public} key.
No public quantum money scheme is provably secure based on standard assumptions.
Some constructions are relative to an oracle, others are without full security proof or are based on post-quantum indistinguishability obfuscation, which is not known to exist based on standard assumptions. 

Franchised quantum money is an intermediate notion. Here, verification uses a franchised key: a unique key that the bank creates for every user in the system. The main drawback is that there needs to be a cap, decided in advance, on the number of franchised keys.

The keyless variant is the most powerful one, yet it is the weirdest: here, there is no master secret key used for minting nor a master verification key. Instead, everyone can create a quantum money state with an associated public key. The unforgeability guarantee is that a quantum polynomial-time adversary cannot create two quantum states that would pass verification with respect to the \emph{same} public key (except with negligible probability). How is that possible? Here we must use the no-rewinding lemon: the minting algorithm uses a measurement for generating the public key. Running the same algorithm twice would yield two different measurement outcomes and post-measured states. 

The term coined for this type of quantum money is quantum lightning, resembling the proverb ``lightning never strikes the same place twice''. 
Quantum lightning also addresses the question of verifiable randomness.
Usually, we think of this public key as a form of a serial number that helps us to validate the quantum money~\cite{FGH+12}.
Zhandry suggested an alternate view~\cite{Zha21}: that the quantum state serves as a proof regarding some property of this serial number, and, more specifically, its randomness. 
Of course, the serial number itself is just a fixed string. 
What is argued is that the process which produced that serial number must generate randomness.
More precisely, consider the process that generated the quantum lightning state and the serial number, and think of the serial number as a random variable. This random variable must have a high min-entropy.
How come? Suppose it had logarithmic min-entropy. 
Then, by running the process polynomially many times, we would end up with two states that pass verification with the \emph{same} serial number, violating the security definition of quantum lightning. 

\paragraph{Classical verifiability, proof of destruction, and uncloneable signatures}
Consider private quantum money. To verify the money, the user has to send the quantum money state for verification to the bank via \emph{quantum communication}. In classically verifiable schemes, this can be achieved by an interactive protocol with only classical communication: usually, the bank sends some challenge message to the user (e.g., specifying in which basis every qubit should be measured~\cite{PYJ+12}), and the user sends an answer to the challenge (in the previous example, the measurement outcomes). The user should not be able to pass strictly more than $n$ verifications given $n$ quantum money states. 
One way of using that is that the user could attach an ID of another user in response to the challenge. Upon successful verification, the bank could mint fresh quantum money and send it to the recipient. 

In other schemes, there is a non-interactive Proof of Destruction (PoD): there is a way for the user to generate a proof that the money has been destroyed. Since PoD is a non-interactive process, some attention is needed: the user might try to claim that the money has been destroyed to several different bank branches. Therefore, this form of Proof of Destruction cannot support multiple non-communicating bank branches (which Wiesner's money definitely can).

In the most advanced form, the quantum state can be used to \emph{sign} an arbitrary message. The unforgeability guarantee is that generating $n+1$ different signed messages is impossible given $n$ such states. In the keyless variant, the adversary's goal is to generate two distinct signed messages associated with the \emph{same} public key.

Here is a motivating example for uncloneable signatures outside financial settings: Suppose Alice goes on vacation and wants Bob to be able to sign a message on her behalf while she is away. If she gives Bob her standard digital signature signing key, Bob would be able to sign an arbitrary number of messages on her behalf. Instead, Alice could generate an uncloneable quantum signing key that would allow Bob to sign at most one (arbitrary) message.

\paragraph{Classical generation} In most of the schemes discussed, the generation of the quantum state is done by the party who holds the secret key. In schemes with classical generation (in the context of money, this is often called classical minting), the generation is an interactive protocol with classical communication between the user and the party holding the secret key. Schemes with classical verifiability and classical generation are called semi-quantum since the party holding the secret key (e.g., the bank) need not have quantum resources.

\paragraph{Indistinguishable copies} In most cases, the quantum state has some classical information attached to it, such as the serial number in Wiesner's scheme. This serial number and distinguishing features of the quantum money state can be used to \emph{trace} users and to violate their privacy. In a scheme with indistinguishable copies, all the quantum states are exact copies of each other, which assures untraceability\onote{add citation}. 
%These are called quantum coins, which, similarly to coins, do not have any distinguishing features, unlike banknotes, which often have a (unique) serial number.
Quantum money with this feature is called a quantum \emph{coins} scheme since coins---unlike banknotes---do not carry a serial number. The techniques for constructing quantum coins are often related to pseudorandomness, see \cpageref{sec:pseudorandomness}\onote{Perhaps, simply, see the following section?}.

% \makesavenoteenv{table*}
% \makesavenoteenv{tabular}
\begin{table*}[ht]
\begin{minipage}{\linewidth}
\caption{\label{tab:quantum_money}Variants of quantum money and uncloneable signatures. The letter q. abbreviates quantum.}

%%% arXiv version
\hspace{-1.5cm} 
\resizebox{1.15\textwidth}{!}{%
\begin{tabularx}{1.27\textwidth}{@{}lccc@{}}

%CACM version
%\resizebox{0.95\textwidth}{!}{%
%\begin{tabularx}{\textwidth}{@{}lccc@{}}

%%%% COPY BELOW THESE LINES

\toprule & (1) Money & (2) Money with Classical Verifiability/PoD & (3) Signatures \\
\midrule (1) Private &  &  &  \\
\hspace{6mm} Standard & ~\cite{Wie83}\tablefootnote{\label{noise tolerant}Noise tolerant.} & classically verifiable q. money~\cite{Gav12}\tablefootnote{Quantum money in this scheme can be verified several times. The classical verification does not destroy all of it.},~\cite{PYJ+12,MVW12}\tablecref{noise tolerant} & q. tokens for MAC~\cite{BS16},~\cite{BSS21}\tablecref{noise tolerant} \\
\hspace{6mm} Classical generation & ? & private semi-q. money~\cite{RS22} & ? \\
\hspace{6mm} Indistinguishable copies & q. coins~\cite{MS10}\tablefootnote{Inefficient construction.},~\cite{JLS18} & ? & ? \\ 
\cdashlinelr{1-4} % THIS LINE NEEDS THE CODE AT THE TOP OF THE PAGE
(2) Franchised &  &  &  \\
\hspace{6mm} Standard & ~\cite{RZ21} & ? & ? \\
\hspace{6mm} Classical generation & ? & ? & ? \\
\hspace{6mm} Indistinguishable copies & ? & ? & ? \\
\cdashlinelr{1-4}
(3) Public &  &  &  \\
\hspace{6mm} Standard & ~\cite{AC13}\tablefootnote{\label{oracle}The construction is relative to a classical oracle.},~\cite{Zha21} & ? & q. tokens for digital signatures ~\cite{BS16}\tablecref{oracle},~\cite{CLLZ21} \\
\hspace{6mm} Classical generation & ? & public semi-q. money~\cite{Shm22} & semi-q. tokens for digital signatures ~\cite{Shm22b} \\
\hspace{6mm} Indistinguishable copies & public q. coins~\cite{BS20}\tablefootnote{The construction has several drawbacks, see ~\cite{BS20} for details.} & ? & ? \\ \cdashlinelr{1-4}
(4) Keyless & % &  &  \\ 
%\hspace{6mm} standard &
q. lightning ~\cite{FGH+12}\tablefootnote{No security proof},\cite{Zha21}\tablefootnote{\label{non-collapsing}Assumes the existence of non-collapsing hash-functions, for which there are no provable constructions, or even candidate constructions.} & q. lightning with bolt-to-certificate ~\cite{CS20}\tablecref{non-collapsing} & one-shot signatures ~\cite{AGKZ20}\tablecref{oracle} \\
%\hspace{6mm} indistinguishable copies & ? & ? & ? \\

%%%% COPY ABOVE THIS LINE
\bottomrule
\end{tabularx}% 
}
%\vspace{-7mm}
\end{minipage}
\end{table*}

\paragraph{The mini-scheme technique}
In a (full) money scheme, an adversary breaks the unforgeability property by passing $n+1$ verifications, given only $n$ quantum banknotes from the bank. A quantum mini-scheme provides a weaker guarantee: the adversary breaks the unforgeability by succeeding in passing $2$ verifications given a single banknote. There are techniques to lift any quantum money mini-scheme to a full scheme, with the additional assumption that post-quantum one-way functions exist. Mini-schemes were first introduced in the context of public quantum money in Ref.~\cite{AC13}, but can also be used in other settings.

The lifting of a public quantum money mini-scheme to a full scheme is achieved as follows: The bank generates a digital signature key-pair as the public quantum money secret key and public key. To generate a banknote, the bank generates a single mini-scheme banknote and signs the mini-scheme's public key. The full scheme verification is done by running the mini-scheme's public verification, and checking that the public key has a valid signature by the bank. 

In the context of uncloneable signatures, there is another simplification. In a full scheme, multiple quantum signing keys can be generated, and each one can sign an arbitrarily long message. The unforgeability guarantee is that an adversary that is given $n$ quantum signing keys cannot produce $n+1$ distinct messages. In a mini-scheme, a single quantum signing key is generated and can be used to sign only one bit. In a mini-scheme, it should be intractable to sign both $0$ and $1$ using a single quantum signing key (except with negligible probability). To support arbitrary long messages, the lifting of a mini-scheme to a full scheme for uncloneable signatures also uses the hash-and-sign paradigm, see ~\cite{BS16,BSS21} for details. 

\paragraph{The hidden subspace technique}
Aaronson and Christiano~\cite{AC13} introduced the subspace hiding technique, which proved useful in several uncloneable primitives. %A prominent technique, which was introduced in Ref.~\cite{AC13}, uses a \emph{subspace state} for the quantum money mini-scheme mentioned. 
%A prominent technique, which was introduced in Ref.~\cite{AC13}, uses a \emph{subspace state} for the quantum money mini-scheme mentioned. 
Suppose $S$ is a random $\frac{n}{2}$ dimensional subspace of $\mathbb F_2^n$. We define the subspace state, denoted $\ket{S}$ as $\frac{1}{\sqrt{|S|} } \sum_{x\in S} \ket{x}$.
%This state turned to be useful in various settings, including as a quantum money mini-scheme, uncloneable signatures, and copy protection.
This state has a number of useful properties:
\begin{itemize}
        \item Given a basis for $S$, the state $\ket{S}$ can be prepared efficiently. This allows the bank to generate a banknote. 
        \item Given oracle access to membership in $S$ and $S^\perp$, the 2-outcome measurement $\{\ketbra{S},I-\ketbra{S}\}$ can be implemented efficiently, where
        %state $\ket{S}$ can be efficiently verified given oracle access to membership in $S$ and $S^\perp$, where 
        \[ S^\perp := \left\{y\in \mathbb F_2^n \mid \forall x\in S,\, %x \cdot y=
        \sum_{i=1}^n x_i y_i = 0 \pmod 2   \right\}\text{.}\]
        This property allows public verification by the users, assuming access to these membership oracles.
        \item Given a single copy of $\ket{S}$, it is impossible to efficiently clone this state, even with oracle access to membership in $S$ and $S^\perp$, except with negligible probability. This property is sufficient to prove unforgeability.
\end{itemize}

The items above were used by Aaronson and Christiano~\cite{AC13} to construct a public quantum money mini-scheme relative to an oracle. Aaronson and Christiano also provided a construction without an oracle, but it was broken in multiple works---see~\cite{PDF+19} and references therein. To recover from that attack, Zhandry~\cite{Zha21} showed that instead of the membership oracles, the bank could use \emph{indistinguishability obfuscation} (iO) to obfuscate these subspace membership functions while preserving the security of the scheme. For more details regarding iO---a cryptographic tool so powerful that it is often referred to as crypto-complete---see~\cite{Bar16}.

The hidden-subspace technique can also be used to construct a public uncloneable signature mini-scheme. 
\begin{itemize}
        \item Given a single copy of $\ket{S}$, one can find a non-zero element of $S$ by measuring it in the standard basis, or a non-zero element of $S^\perp$ by measuring $\ket{S}$ in the Hadamard basis. Here, $\ket{S}$ serves as the quantum signing key, a non-zero element in $S$ serves as a signature of $0$, and a non-zero element of $S^\perp$ serves as a signature of $1$. The signatures could be easily verified using the subspace membership oracles.
        \item Given a single copy of $\ket{S}$, it is impossible to find a non-zero element of $S$ \emph{and} a non-zero element of $S^\perp$, even with oracle access to membership in $S$ and $S^\perp$, except with negligible probability. This property is sufficient to prove the unforgeability of the uncloneable signature scheme. 
        %Yet, one cannot find a non-zero element of $S$ \emph{and} a non-zero element of $S^\perp$, using polynomially many queries to the membership oracles mentioned above except with non-negligible probability.
\end{itemize}
For more details regarding the items above, see Ben-David and Sattath~\cite{BS16}. Coladangelo et al.~\cite{CLLZ21} also showed how the oracles could be removed by obfuscating the subspace membership functions.\footnote{They also change the scheme slightly: for technical reasons which are outside the scope of this work, they use a novel related notion called coset states instead of the subspace states that were introduced earlier.}
The hidden-subspace technique was also used in quantum copy-protection.

\subsection*{Quantum pseudorandomness}
\label{sec:pseudorandomness}

Pseudorandomness plays a crucial role in cryptography, as well as in theory of computer science. 
Pseudorandom states (PRS) is a family of efficiently generated quantum states $\{\ket{\psi_k}\}_{k\in \{0,1\}^m}$ such that it is computationally hard to distinguish between polynomially many copies of: (a) $\ket{\psi_k}$ sampled uniformly from the family, and (b) a uniformly (Haar) random quantum state~\cite{JLS18}. 

The notion of a PRS family is a variant of quantum state t-designs~\cite{AE07}: $t$-designs offer statistical rather than mere computational indistinguishability but have the drawback that they are only secure in the presence of $t$ copies, where $t$ is predetermined. Pseudorandom states remain computationally secure regardless of the number of states produced.

The simplest construction of pseudorandom states is the following. Let $PRF:\mathcal K\times \mathcal X \to \{0,1\}$ be family of pseudorandom functions. Let $\ket{\psi_k}:=\frac{1}{\sqrt{|\mathcal X|}} \sum_{x\in \mathcal X} (-1)^{PRF_k(x)} \ket{x}$. Brakerski and Shumeli~\cite{BS19} proved that the family $\{\ket{\psi_k}\}_{k \in \mathcal K}$ is a PRS family. This proved a conjecture by Ji, Liu and Song~\cite{JLS18}, and simplified their earlier construction.

% This notion is a weakening of a quantum t-design~\cite{AE07} in that it weakens statistical indistinguishability with a computational one, and strengthens it in the sense that it holds for any polynomial number of copies, rather than $t$ copies. 

Kretschmer~\cite{Kre21} proved that, to some extent, pseudorandom states are \emph{weaker} than one-way functions, arguably the minimal assumption in classical cryptography: he showed that there is no black-box reduction from pseudorandom states to one-way functions. Note that the construction mentioned in the previous paragraph is a black-box reduction in the opposite direction.
Kretschmer's result raises the possibility of basing quantum cryptography on pseudorandom states---an assumption that may be weaker than the existence of one-way functions.
%Indeed, recent results~\cite{MY21,AQY21} show how pseudorandom quantum states imply a computationally hiding and statistically binding quantum commitment scheme for classical information. In the classical setting, such a commitment scheme exists if and only if one-way functions exist. 
Indeed, several recent results show how to construct basic cryptographic tasks based on PRS (or its variants). Notable examples are: a computationally hiding and statistically binding quantum commitment scheme~\cite{MY21,AQY21}, quantum multi-party computation for classical circuits~\cite{AQY21,MY21}, symmetric encryption with quantum ciphers~\cite{AQY21}, private quantum coins~\cite{JLS18}, almost public quantum coin~\cite{BS20} and Lamport (i.e., one-time secure) signatures with quantum public keys~\cite{MY21}. A remaining open question is whether digital signatures are implied by pseudo-random states.

We mention briefly that a similar approach can be used to define pseudorandom \emph{unitaries} and \emph{unitary} t-designs. Pseudorandom unitaries and unitary $t$-designs can be used to generate a state $t$-design, but not conversely. There are still no constructions for pseudorandom unitaries, though Ji, Liu and Song~\cite{JLS18} suggest two constructions that they conjecture to be secure. Their first construction is based on pseudorandom functions, and another is based on pseudorandom permutations.

\cref{tab:pseudorandomness} presents the classical analogs of the quantum pseudorandom objects discussed above. 
%The notions of quantum pseudorandom objects discussed above have classical analogies. See \cref{tab:pseudorandomness} for details. 

% \setlength{\tabcolsep}{3pt}
%\rowcolors{2}{white}{gray!25}

\begin{table}[ht]
{\renewcommand{\arraystretch}{1.5}
\caption{\label{tab:pseudorandomness} Classical and quantum forms of pseudorandomness.}

%%%%% CACM version
%\begin{tabular*}{\columnwidth}{lcc}
%%%%% arXiv version
\begin{tabular}{lcc}
%%%% COPY BELOW THESE LINES

\toprule & Statistical guarantees & Computational guarantees \\ 
 \midrule \multicolumn{3}{c}{Random configurations} \\
\addlinespace[0.1cm] classical & \makecell[l]{t-wise independent \\ distributions, folklore}    & pseudorandom generators~\cite{BM84} \\
quantum                        & state t-designs~\cite{AE07}                                    & pseudorandom states~\cite{JLS18} \\
\cdashlinelr{1-3} \multicolumn{3}{c}{Random transformations} \\
\addlinespace[0.1cm] classical & \makecell[l]{t-wise independent \\ hash functions~\cite{CW79}} & pseudorandom functions~\cite{GGM86} \\
quantum                        & unitary t-designs~\cite{DCE+09}                                & pseudorandom unitaries~\cite{JLS18}

%%%% COPY above THESE LINES

\\ \bottomrule 
%%%% CACM version:
%\end{tabular*}
%%%% arXiv version:
\end{tabular}%

}
\end{table}

\onote{Discuss applications}

\subsection*{Quantum copy-protection}
Software companies want to sell programs that can be used offline by the customers (even if the installation requires online communication). On the other hand, they want to restrict users from sharing the programs with others.  
Classically, this restriction cannot be enforced: a pirate could always create a perfect copy of the computer with the installed program, including all its component (the hard disk drive, CPU, memory, etc.), and at this point, both computers would have a working copy of the program.

Quantum copy protection is a compiler that receives a classical program (for example, the source code, or a Boolean circuit) as the input, and returns a copy-protected program in the form of a quantum state. This copy-protected program can be used to run the program on any input, as many times as the user wishes.

The security guarantee is the following. A challenger samples a program $P$ from a distribution over programs $\mathcal{P}$. The pirate is given the copy-protected state of the program $P$. The pirate uses this state and sends two states to two freeloaders. Then, a random challenge $x$ is sampled and handed to both freeloaders, and each freeloader sends back $y_i$, which is their attempt to compute $P(x)$. Note that the freeloaders cannot communicate at this point. They win if $y_0=y_1=P(x)$. Note that one freeloader could easily compute $P(x)$ correctly: the pirate could send the copy-protected program to the first freeloader, which the freeloader could use to evaluate $P(x)$.
The seconder freeloader would have to guess $P(x)$; therefore, the chances could be slim for certain distributions $\mathcal{P}$.  

A distribution over programs $\mathcal P$ is \emph{quantum learnable} if the program can be learned efficiently from its \onote{by?} input-output behavior with a quantum computer. 
%More precisely, $\mathcal P$ is quantum learnable if there exists a quantum polynomial-time algorithm $\mathcal L$ with quantum oracle access to $P$, where $P$ is sampled from $\mathcal P$. When the learning phase is over, the oracle access is revoked, and the learner gets a random $x$. We say that the learning succeeds if $L$ can compute $P(x)$. 
It is impossible to copy-protect a learnable distribution: the pirate could learn the program using one copy of the copy-protected program and send the insights from the learning phase to the freeloaders.

This security definition can be easily strengthened so that there are $n$ copy-protected programs and $n+1$ freeloaders, and cases where the challenge input $x$ is not necessarily uniformly random.
\onote{Ok, here is another try. I changed "quantum learnable" to be that you learn the entire function. Clearly, these cannot be copy-protected.}

There are constructions for quantum copy-protection for any unlearnable family relative to quantum~\cite{Aar09} and classical~\cite{ALL+21} oracles, and no construction is known in the plain model for any class. 
\snote{There is "semi-honest" copy protection (two freeloaders, one honest and one malicious) with no assumptions.} 
Ananth and La Placa~\cite{AP21} introduced a weaker form called secure software leasing (SSL), which allows for \emph{copy-detection}, and showed a construction for some unlearnable family. Their construction completely reveals the software, but a pirated copy can be \emph{detected} by honest users. On the negative side, they constructed a quantum unlearnable family that cannot be SSLed (and hence, cannot be copy-protected as well).

To this day, copy protection of \emph{quantum} circuits has not been studied. 

\paragraph{One-time programs} One may consider a variant of quantum copy-protection in which the user can evaluate the program only once\cite{BGS13}. One could hope that\snote{I would replace with "One-timeness follows from" or something to that effect.} measuring the output would collapse the quantum state, causing it to self-destruct, similarly to uncloneable signatures. Unfortunately, one-time programs are impossible to construct, even with computational assumptions~\cite{Aar09}. Why is that? Suppose the evaluation gives the correct answer without any errors (i.e., with probability $1$). In this case, the post-measured state remains exactly the same prior to the measurement and therefore does not self-destruct. It can be shown based on the "Almost good as new lemma" that if the correct outcome occurs with probability $1-\epsilon$, the trace distance between the pre-measured state and the post-measured state is $1-O(\sqrt{\epsilon})$, see \cite[Lemma 4]{Aar09}, and therefore can be ignored as long as $\epsilon$ is negligible.  
Suppose errors occur with constant error, say, $\frac{1}{3}$. In that case, the adversary could use $k$ copies of the program to create an enhanced program, in which the error happens with probability $2^{-\Omega(k)}$ by standard error-reduction techniques. 
This technique allows a user with $k$ copies to evaluate the function on a number of points that is exponential in $k$. 

Broadbent, Gutoski and Stebila~\cite{BGS13} showed how to construct quantum one-time programs in a (non-standard) model in which one-time classical memory exists. A \emph{one-time memory} is a box that can store two bits. The user can ask the box for one of the bits and receive the answer, but at this point, the box self-destructs, and the user has no way of retrieving the other bit. One-time memory is an idealized, non-interactive version of 1-of-2 oblivious transfer. A construction of one-time classical programs was already known based on the same assumption~\cite{GKR08}. Refs.~\cite{BGZ21,CGL19} show how to construct one-time programs in other non-standard models. 

\subsection*{Uncloneable forms of encryption}
\paragraph{Uncloneable encryption}
In classical encryption schemes, an eavesdropper can always copy the ciphertext. 
In uncloneable quantum encryption, a classical plaintext is encrypted to a quantum ciphertext. 
The security guarantee is that there is no way for an eavesdropper to manipulate the ciphertext and send it to two isolated parties so that both parties could correctly decrypt it, even after they are provided with the secret key. 

Here is a motivating example for uncloneable encryption. Consider an adversary who can somehow break the encryption scheme and learn the secret key for the ciphertext, but this process is slow. One such instance is an adversary who expects to have a quantum computer (which we know can break most contemporary public-key encryption schemes) in the coming years. 
With an uncloneable encryption scheme, this approach becomes moot since the adversary cannot copy the ciphertext without being detected: if the eavesdropper tries to keep a state which would be enough to recover the plaintext at a later time, this necessarily means that the intended recipient, which has the secret key, would not be able to decrypt the original message that was sent.

Tamper-evident quantum encryption is a weaker variant first introduced by Gottesman~\cite{Got03}: tamper-evident encryption is to uncloneable encryption what SSL is to copy-protection. Several definitions and constructions in the spirit of the informal one given above were first shown by Broadbent and Lord~\cite{BL20} for symmetric uncloneable encryption. Ananth and Kaleoglu~\cite{AK21} have extended this notion to the asymmetric setting. Furthermore, they showed that a strong uncloneable encryption variant implies a weak copy-protection form for a family of point functions. 

\paragraph{Quantum encryption with certified deletion}
Alice has a terminal illness and wants Bob to have all her passwords once she passes away.
She encrypts her message and sends the ciphertext to Bob, and the decryption key to her lawyer, who is ordered to send the secret key to Bob upon her death.

Suddenly, a cure for Alice's illness is found, and she wants to ensure her passwords will not leak.
Even if Bob complies, there is no way to certify that he had not saved a copy unbeknownst to Alice.
Quantum encryption with certified deletion is an encryption scheme of classical messages with quantum ciphertexts that, in addition to a standard privacy guarantee, allows Bob to prove that he deleted the ciphertext and would not be able to decrypt the passwords even should he learn the decryption key~\cite{BI20} (see also~\cite{HMNY21}). More precisely, an interactive protocol with classical communication can be used to revoke Bob's ciphertext. A successful revocation guarantees that even if Bob gets the decryption key after the revocation, he would not be able to decrypt the passwords. 

\paragraph{Revocable quantum timed-release encryption}
Suppose Alice wants to send Bob a message that he could read only one year ahead.
In a (classical) time-release encryption, the encryption is efficient, but decrypting the ciphertext requires a number of computational steps that Alice can choose upon encryption. Assuming Alice can estimate Bob's computer clock rate, she could set the encryption parameters so that Bob could recover her message from the ciphertext a year ahead. 

However, suppose, as in the example with Alice's terminal illness, a cure is found after only one month. As was already discussed in the context of quantum encryption with certified deletion, \emph{revoking} a classical encryption in this setting is impossible.
In a revocable quantum timed-release encryption~\cite{Unr15}, the ciphertext is quantum, and there is a protocol that allows Bob to prove that he deleted the quantum ciphertext before the time-release deadline arrives---i.e., the one year mentioned in the example above. The main advantage of using quantum timed-release encryption (over quantum encryption with certified deletion) is that there is no need to use a trusted party---the lawyer in the earlier example.

\paragraph{Uncloneable decryptors}
Subscription-based satellite television works roughly as follows: the TV shows are encrypted, and the satellite broadcasts the shows in encrypted form, which can be received by \emph{anyone} with a satellite dish. Paying customers receive a smart-card containing a decryption key that allows them to view the television shows. The problem? A sophisticated pirate can create many copies of the smart-card. This attack is not merely a theoretical risk: see \url{https://en.wikipedia.org/wiki/Conditional_access} for a list containing systems known to be compromised.

A (symmetric or asymmetric) encryption scheme with uncloneable decryptors is a scheme that can be used to produce \emph{uncloneable} quantum decryption keys~\cite{GZ20,SW22}. The decryptor can be used to decrypt the (classical) ciphertexts: The decryptor allows the subscribers to watch the TV shows in the example above. The uncloneability feature guarantees that an adversary that receives a decryptor state should not be able to produce two states given to two distinguishers, that could use it to win the standard semantic security indistinguishability game. This notion can be generalized to support $k$ decryptors and $\ell>k$ distinguishers.  

\onote{more open questions?}
\section*{Discussion}
Quantum key distribution (QKD) has been commercially available for two decades; quantum computers are being developed by several industry giants hoping to demonstrate quantum advantage with real impact soon. 

Uncloneable cryptography, for the most part, makes little sense without long-term quantum memory. For example, one of the functions of money is as a \emph{store of value}. Quantum money, which could not be stored even for a second---more than the lifetime of the qubits in all the major quantum computing platforms being developed---would not be particularly useful. It remains to be seen whether uncloneable cryptography would have applications in the coming, noisy intermediate-scale quantum (NISQ) era. 
% Prior to the groundbreaking idea of Merkle, Diffie and Hellman, public-key cryptography was hard even to imagine. Perhaps we are facing a similar challenge, as it is hard to imagine other applications for uncloneability. 

\begin{acks}
This work was supported by the Israel Science Foundation (ISF) grants No. 682/18 and 2137/19 and
by the Cyber Security Research Center at Ben-Gurion University. We thank Dean Doron for valuable discussions regarding various notions of pseudorandomness, Shai Wyborski for his comments, and Vince Serrano for his illustration in \cref{fig:lemons}.
\end{acks}

\bibliographystyle{ACM-Reference-Format}
\bibliography{unclonable_cryptography}

% \appendix


\end{document}